\documentclass[sigconf, nonacm, 9pt]{acmart}

\usepackage{amsmath,amsfonts}

\usepackage[vlined,ruled,linesnumbered]{algorithm2e}

\usepackage[font=small,skip=10pt]{caption}
\usepackage{caption,subcaption}
\usepackage{dblfloatfix}
\usepackage{multirow}
\usepackage{paralist}
\usepackage{cprotect}
\usepackage{dirtytalk}
\usepackage{amsmath}
\usepackage{booktabs} 
\usepackage{graphicx}
\usepackage{svg}
\usepackage{xcolor}
\usepackage{siunitx}
\usepackage{todonotes}
\usepackage{pgfplots}
\usepackage{pgfpie3}
\usepackage{bchart}
\usepackage[eulergreek]{sansmath}
\usepackage{helvet}
\usetikzlibrary{arrows}

\definecolor{mpink}{HTML}{ef476f}
\definecolor{myellow}{HTML}{ffd166}
\definecolor{mgreen}{HTML}{06d6a0}
\definecolor{mblue}{HTML}{118ab2}
\definecolor{mdarkblue}{HTML}{073b4c}

\definecolor{ared}{HTML}{ff898d}
\definecolor{ayellow}{HTML}{ffda74}
\definecolor{ablue}{HTML}{afe160}
\definecolor{agreen}{HTML}{4bace8}
\definecolor{apurple}{HTML}{967bbb}

\definecolor{modern_green}{HTML}{70AD47}
\definecolor{modern_blue}{HTML}{00B0F0}
\definecolor{modern_lightblue}{HTML}{A1DFF6}
\definecolor{modern_yellow}{HTML}{FFC000}
\definecolor{modern_red}{HTML}{E42831}
\definecolor{modern_darkgray}{HTML}{434343}
\definecolor{modern_darkblue}{HTML}{0083b2}
\definecolor{modern_lred}{HTML}{F2999D}
\definecolor{modern_lgreen}{HTML}{c5e384}
\definecolor{modern_lyellow}{HTML}{fce883}
\definecolor{modern_violet}{HTML}{967bbb}

\definecolor{moderngreen}{HTML}{70AD47}
\definecolor{modernblue}{HTML}{00B0F0}
\definecolor{modernlightblue}{HTML}{A1DFF6}
\definecolor{modernyellow}{HTML}{FFC000}
\definecolor{modernred}{HTML}{E42831}
\definecolor{moderndarkgray}{HTML}{434343}
\definecolor{moderndarkblue}{HTML}{0083b2}
\definecolor{modernlred}{HTML}{F2999D}
\definecolor{modernlgreen}{HTML}{c5e384}
\definecolor{modernlyellow}{HTML}{fce883}
\definecolor{modernviolet}{HTML}{967bbb}

\definecolor{blue1}{HTML}{03045e}
\definecolor{blue2}{HTML}{023e8a}
\definecolor{blue3}{HTML}{0077b6}
\definecolor{blue4}{HTML}{0096c7}
\definecolor{blue5}{HTML}{00b4d8}
\definecolor{blue6}{HTML}{48cae4}
\definecolor{blue7}{HTML}{90e0ef}
\definecolor{blue8}{HTML}{ade8f4}
\definecolor{blue9}{HTML}{caf0f8}

\renewcommand{\sfdefault}{phv} 
\pgfplotsset{
  compat =1.17,
  tick label style = {font=\sansmath\sffamily\scriptsize},
  every axis label = {font=\sansmath\sffamily\scriptsize},
  legend style = {font=\sansmath\sffamily\scriptsize},
  label style = {font=\sansmath\sffamily\scriptsize},
}

\newcommand\eat[1]{}

\newcommand\notes[1]{\textcolor{red}{#1}}
\newcommand\notesdone[1]{\textcolor{black}{}}

\graphicspath{{images/}}

\newcommand{\fw}{{AMPER}\xspace}
\newcommand{\csp}{CSP\xspace}
\newcommand{\fnn}{{frNN}\xspace}
\newcommand{\fwk}{{AMPER-k}\xspace}
\newcommand{\fwf}{{AMPER-fr}\xspace}

\begin{document}

\title[AMPER for Deep RL]{Associative Memory Based Experience Replay \\for Deep Reinforcement Learning}

\author{Mengyuan Li, Arman Kazemi, Ann Franchesca Laguna and X. Sharon Hu} 
\email{{mli22, akazemi, alaguna, shu}@nd.edu}
\affiliation{
  \institution{Department of Computer Science and Engineering, University of Notre Dame}
  \city{Notre Dame}
  \state{Indiana}
  \country{USA}
  \postcode{46556}
}

\renewcommand{\shortauthors}{M. Li, et al.}

\begin{abstract}
 
Experience replay is an essential component in deep reinforcement learning (DRL), which stores the experiences and generates experiences for the agent to learn in real time. Recently, prioritized experience replay (PER) has been proven to be powerful and widely deployed in DRL agents. However, implementing PER on traditional CPU or GPU architectures incurs significant latency overhead due to its frequent and irregular memory accesses. This paper proposes a hardware-software co-design approach to design an associative memory (AM) based PER, \fw, with an AM-friendly priority sampling operation. \fw replaces the widely-used time-costly tree-traversal-based priority sampling in PER while preserving the learning performance. Further, we design an in-memory computing hardware architecture based on AM to support \fw by leveraging parallel in-memory search operations. \fw shows comparable learning performance while achieving 55$\times$ to 270$\times$ latency improvement when running on the proposed hardware compared to the state-of-the-art PER running on GPU. 
\end{abstract}

\maketitle

\section{Introduction}
\label{sec:introduction}


Deep reinforcement learning (DRL) combining reinforcement learning and deep learning is a powerful framework for agents to learn to make decisions based on trial and error. DRL can be used in many applications such as gaming, robotics and other automated systems~\cite{haarnoja2019learning}. Some DRL methods learn offline, while others conduct learning online where an agent learns as it interacts with the environment. Online DRL is preferred when the environment is complex and changes often. It is highly desirable for Online DRL to satisfy certain real-time latency constraints. Deep Q-network (DQN), first introduced by Google DeepMind in~\cite{mnih2013playing}, is a popular, model-free, online, off-policy DRL method. 

In DQN, an agent learns through past experiences which are described by state transitions, rewards, and actions. A DQN agent is comprised of three main components: (1) an \textit{action network} which determines the action at each time step for a given input state, (2) a \textit{target network} which learns from past experiences, and (3) an \textit{experience replay (ER) memory} which stores experiences and generates specific experiences for the target network as training input. The structure of target and action networks can be multilayer perceptrons (MLPs) or convolution neural networks (CNNs)~\cite{mnih2013playing}. 

For complex environments with many states, the ER memory can be very large, and it can take a significant amount of time to update the memory and generate new experiences. Through detailed profiling of several open-source DQNs running on GPU, we find that ER operation (sampling experiences) can take more than 55\% of the total DQN execution time. Though there is an abundance of work on accelerating neural networks used in the action/target network, few works have considered accelerating the ER related operations. In order to meet real-time latency requirements for online deployment of DQNs, it is critical to devise techniques to accelerate the ER operations in DQN, which is our focus.



The key operation supported by ER memory is sampling a small subset of the stored experiences as the training data for the target network at each time step. ER memory can be very large (e.g. on the order of $10^6$ entries) since the experiences at many past time steps may need to be stored. Hence sampling experiences faces the memory-wall~\cite{han2016eie} challenge for CPU and GPU implementations. Also, sampling techniques involve non-trivial calculations and can significantly impact the learning performance and speed. Uniform sampling was used in the earlier DQNs but its performance was not high. Prioritized experience replay (PER)~\cite{schaul2015prioritized}, deploying priority sampling technique, is widely used in the state-of-the-art DQN implementations like Rainbow~\cite{hessel2018rainbow} and Agent57~\cite{badia2020agent57}. ~\cite{hessel2018rainbow} shows that without PER, the learning score of a DQN agent may drop around 50\%. However, PER requires even more frequent and irregular access to the ER memory, further exacerbating the memory-wall challenge.

In-memory computing, where computation is performed directly inside the memory array, is an effective computing paradigm for addressing the memory-wall challenge~\cite{ielmini2018memory}. Associative memory (AM), a.k.a. content addressable memory (CAM), is an in-memory computing primitive that supports parallel search. AM can reduce the search time from $O(n)$ to $O(1)$ where $n$ is the number of elements to be sought from. 
However, straightforward use of AM does not offer significant gains for PER since the basic tree-traversal steps for priority sampling are sparse and irregular. Hence, using hardware-software co-design, we design an AM based PER algorithm, \fw and an AM based accelerator for \fw. To the best of our knowledge, our proposed method is the first work that targets accelerating ER operations. 
%
%

We specifically make the following contributions: (i) We investigate the DQN execution latency distribution under different ER memory and environment settings and identify that ER operations, especially the priority sampling process, are bottlenecks for implementing a low-latency DRL agent.
(ii) We propose a novel AM-based prioritized experience replay (\fw) algorithm with AM-friendly priority sampling operations, which replace the widely-used time-costly tree-traversal-based priority sampling in PER with TCAM searches, while preserving the learning performance. 
(iii) We propose two variants of \fw, \fwk and \fwf, using two AM-based nearest neighbor search operations: k-Nearest Neighbor and fixed-radius Nearest Neighbor, respectively, to trade off learning performance and latency.
(iv) We design an AM-based in-memory computing hardware architecture by employing ternary CAMs (TCAMs)~\cite{hu2021memory} to accelerate the \fw algorithm. We devise a prefix-based query strategy to approximate fixed-radius Nearest Neighbor search with only a single low-latency TCAM search.
(v) We evaluate \fw on widely used OpenAI gym environments~\cite{1606.01540}. Our results show that \fw achieves comparable learning performance as the PER algorithm. Our evaluations based on circuit-level simulations show that \fw running on the AM-based in-memory computing hardware can achieve up to 270$\times$ latency improvement over PER running on GPU.
 
\section{Background and Motivation}
Below we first present the basics of DQN and PER. We then briefly review the related work on DQN acceleration and AMs, especially TCAMs. Finally, we compare different existing ER techniques, and present the profiling data for a typical DQN implementing different ER techniques (uniform ER and PER) to further illustrate the latency performance characteristics.
\label{sec:background}

\subsection{DQN and Prioritized Experience Replay}\label{sec:PER}

\begin{figure}[t]
    \centering
    \includegraphics[width = 0.85\linewidth]{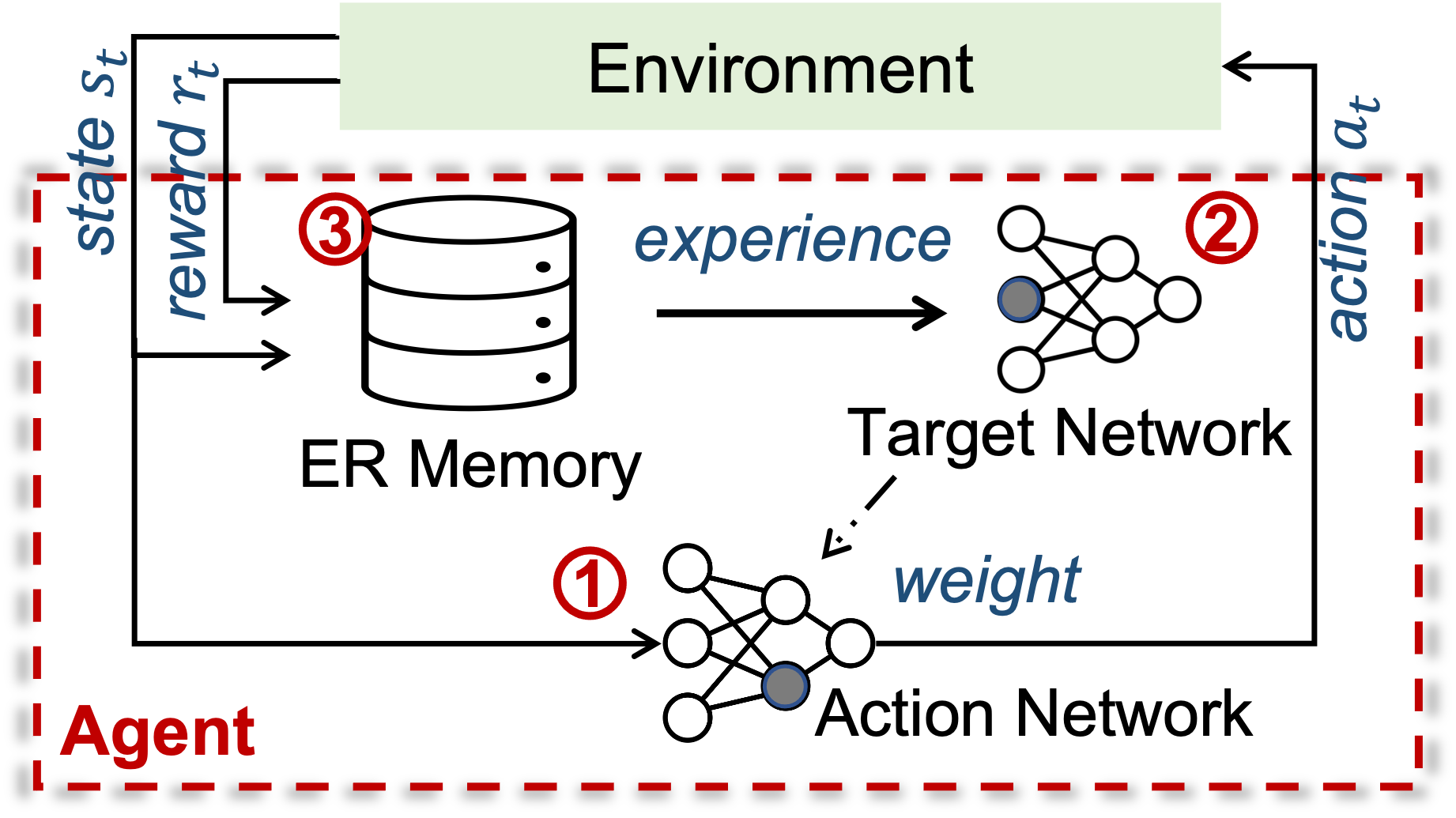}
    \caption{Illustration of a DQN agent interacting with the environment. The agent has three main components: (1) action network, (2) target network, and (3) ER memory.}
    \label{fig:DQN}
    \vspace{-2ex}
\end{figure}

DQN is a model-free, off-policy (i.e., using separate learning and action networks) DRL method which learns through past experiences~\cite{mnih2013playing}. Fig.~\ref{fig:DQN} illustrates a typical DQN agent. At every time step $t$, the agent decides the action $a_t$ via the action network, and uses that action to interact with the environment. The environment then transitions to a new state $s_t$ and generates a reward $r_t$. At each time step, the state transition, the reward, and the action form an experience are stored in ER memory. A random batch of stored experiences are sampled at each time step and fed to the target network to train the agent. The agent learns from the state transitions and rewards by maximizing the global return, defined as the accumulated rewards from the start to the end. 

Experience sampling plays an important role in the learning process. 
Prioritized experience replay (PER), as the state-of-the-art ER technique, frequently samples state transitions that lead to larger reward value change. PER has been empirically shown to improve the performance of the agent compared to the uniform ER~\cite{schaul2015prioritized} which samples the past distribution randomly following a uniform distribution. In PER, the priority sampling technique is deployed where each experience $e_{i}$ is associated with a priority $p_{i}$ determined by the relative magnitude of the temporal-difference error (TD-error). The probability that experience $e_i$ is sampled is defined as $P(i) = \frac{p_i^{\alpha}}{\sum_k p_k^{\alpha}}$
where $p_i$ > 0. The exponent $\alpha$ determines how much prioritization is used, with $\alpha=0$ corresponding to the uniform case. Also, PER needs to update the priority value of each sampled experience with a new TD-error after training is done.

\begin{figure}[t]
    \centering
    \includegraphics[width =0.9\linewidth]{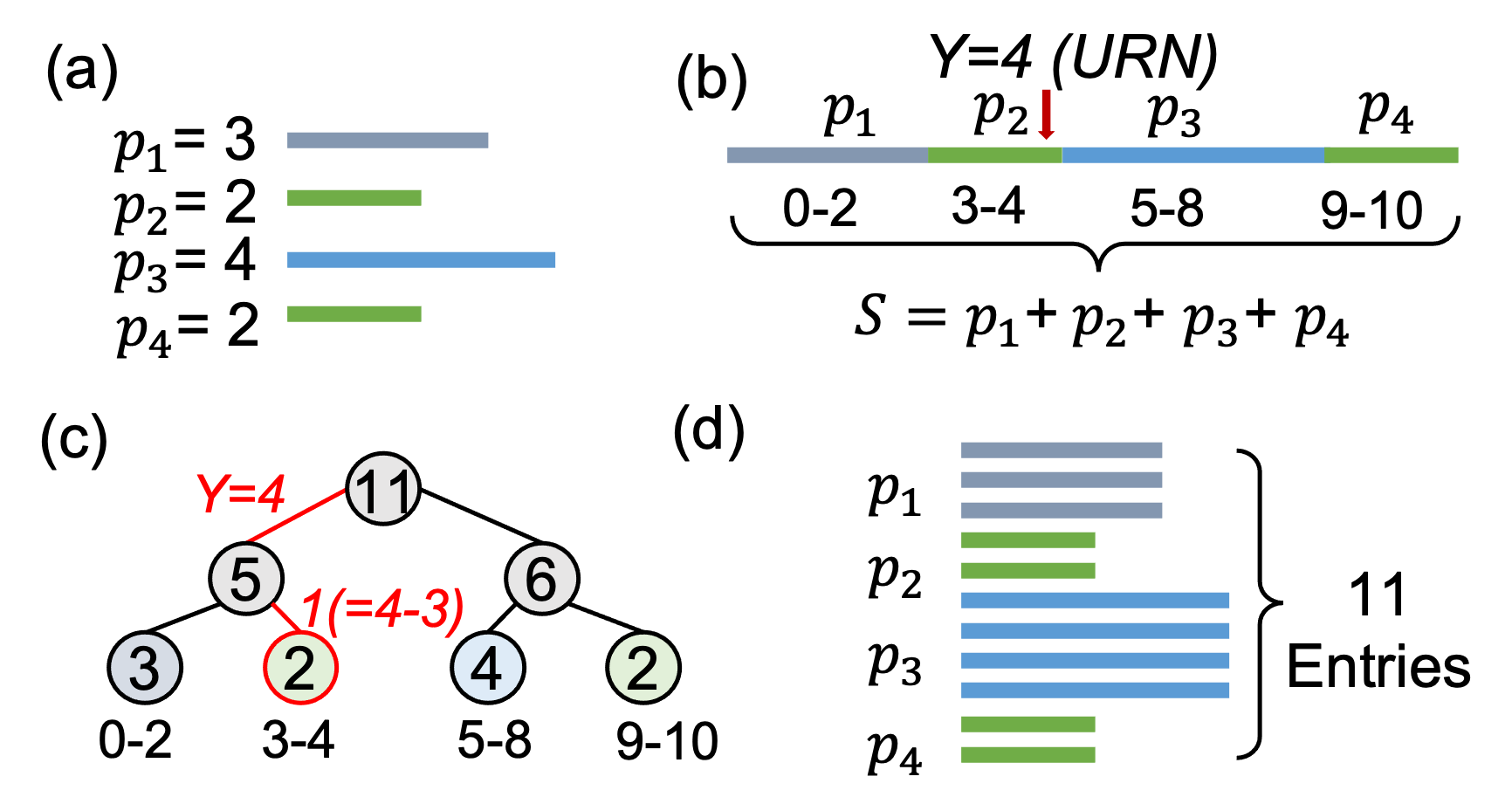}
    \caption{Illustration of PER implementation. (a) An example with 4 prioritized experiences. (b) The basic idea of sum-based sampling. (c) The sum-tree based implementation of (b). Leaf nodes contain the priority values. The search process of $Y=4$ is highlighted in red. (d) A high-level conceptual view of \fw for the example in (a).}
    \label{fig:PERIdea}
    \vspace{-2ex}
\end{figure}

Sampling experiences with PER in a large ERM can be very expensive. A sum-based method is widely used for the priority sampling and is adopted in PER. We illustrate the method in Fig.~\ref{fig:PERIdea}(b) using a simple example with 4 experiences as specified in Fig.~\ref{fig:PERIdea}(a). The sum of the four priorities is $S=p_{1}+p_{2}+p_{3}+p_{4}=11$. A uniform random number (URN) $Y$ is generated from the range $[0, S-1]$. Then, the sampled priority is the one corresponding to the region that $Y$ falls into (e.g., $Y=4$ falls in $p_{2}$ in Fig.~\ref{fig:PERIdea}(b) so the sampled priority is $p_{2}$). It is easy to see that the probability that $Y$ falls into the region of $p_{2}$ is $P(2)=p_{2}/S$. Therefore, by using the sum-based representation,  priority sampling is transformed into uniform sampling without knowing the data distribution. 
The sum-based method is typically realized with a data structure, sum tree, as shown in Fig.~\ref{fig:PERIdea}(c) where sampling is done by search on the sum tree structure. Also, the sum tree is updated when the priority value (leaf node) is updated.
Thus, frequent updates and sampling operations in the DQN learning process incur many tree operations which require frequent access to memory and exhibit irregular memory access patterns, and cause longer latency.

\subsection{Hardware Accelerators for DQN}

Here we briefly review some representative previous work on DQN acceleration. An FPGA based accelerator is proposed in~\cite{su2017neural}. It focuses on accelerating the training and inference of the action and target network, and considers a small ER memory implementing the uniform sampling technique. Some other papers aim to accelerate distributed DQN, where multiple DQN agents work in a distributed fashion. For example, \cite{wang2020many} proposes a customized network-on-chip design to solve the communication problems among the distributed agents. \cite{li2019accelerating} exploits in-switch acceleration to reduce the network communication for gradient accumulation. Previous work usually assumes a small ER memory  and ignores its acceleration. However, for the state-of-the-art DRL agents, a large ER memory is often needed and can incur long latency (more will be shown in Sec.~\ref{sec:breakdown}).

\subsection{In-memory computing and AM}\label{sec:am}

Instead of moving data to the processing unit as in typical von Neumann machines, in-memory computing~\cite{sebastian2020memory} performs computation directly inside the memory in order to solve the memory-wall~\cite{han2016eie} problem. Associative memories (AMs), also known as content addressable memory (CAMs), are in-memory-computing fabrics that support fast and energy efficient search. The two main operations of AMs are (1) search, where the address of the memory entry that matches the input query is identified and (2) write, where data entries are stored in the AM rows. AMs enable parallel searches of a given query against all data stored in memory in $O(1)$ time~\cite{hu2021memory}.

\begin{figure}[t]
    \centering
    \includegraphics[width = \linewidth]{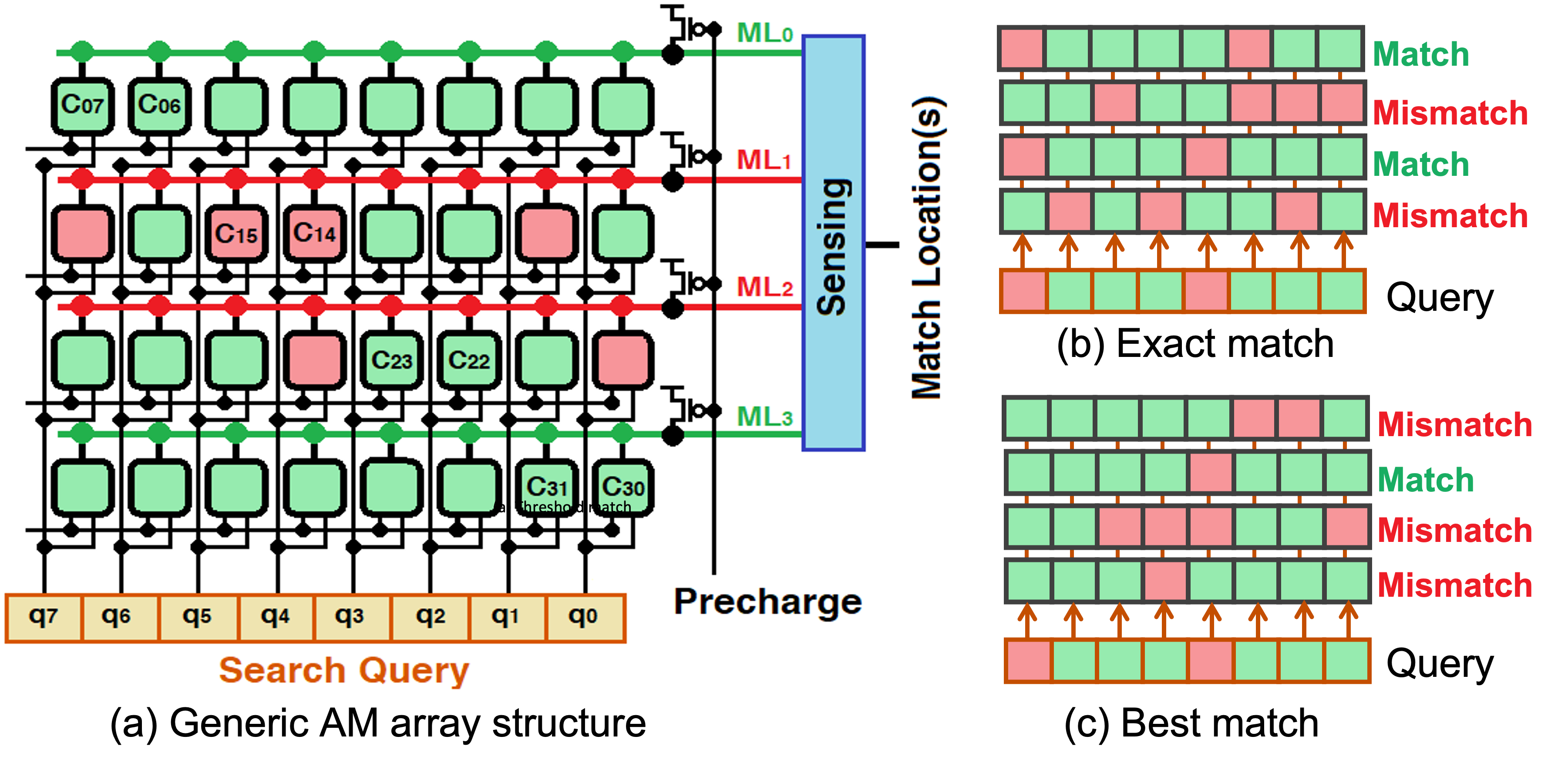}
    \caption{(a) Generic AM array structure (4×8 array) based on the NOR connection. Different match schemes: (b) exact match: the rows that are same as the input query; (c) best match: the row which has the shortest distance from input query is the best match.  \cite{hu2021memory}}
    \label{fig:CAM}
    \vspace{-3ex}
\end{figure}

The most commonly used AM is a Ternary CAM (TCAM) where each element of queries and stored data can assume one of three states: 0, 1, and don't care ('x'). 'x' is a wildcard state which matches with both '0' and '1'. For a TCAM array with $r$ rows and $c$ columns (Fig.~\ref{fig:CAM}(a)~\cite{hu2021memory}), all cells in a row are connected to a common matchline (ML) and each cell stores $C_{ij}$. During the search operation, each cell $C_{ij}$ in row i performs an XNOR operation between its content and the query element $q_j$. If $C_{ij} = q_{j}$, $C_{ij}$ matches the input query (denoted by green), and otherwise there is a mismatch (denoted by red). Each ML implements a logic OR operation of all the cells in the row to determine the result for that row.

Different sensing circuits can be designed to realize different match schemes. One typical match scheme is the exact match as shown in Fig.~\ref{fig:CAM}(b), which reports rows that ``exactly match'' the query for every single cell. Exact match search is the fastest search type due to its simple sensing requirement~\cite{hu2021memory}. Another match scheme is best match, which reports the row with the least number of mis-matching cells. For best-match search, the discharge rate of the ML is proportional to the number of mis-match cells on the ML. Best match (Fig.~\ref{fig:CAM}(c)) search is widely used for nearest neighbor search ~\cite{ni2019ferroelectric}. To find the best match, it is possible to use analog-digital-converters to digitize the the ML voltage~\cite{karunaratne2020memory}, which is a costly approach. Another approach is to use a winner-take-all circuit to find the row with the highest voltage (lowest discharge)~\cite{imani2019searchd}. This approach is more energy and area efficient than using analog-digital-converters but can be limited to finding best matches only within a certain 
number of mis-match cells. In this work, we will exploit the different match schemes to accelerate \fw.

As discussed in Sec.~\ref{sec:PER}, a tree-based method is employed to implement PER. Recent work~\cite{pedretti2021tree} proposed to use AMs to accelerate a tree structure by mapping each path from the root node to a leaf node to one row of the AM. This kind of mapping can indeed accelerate the search process, but it exhibits poor latency performance for update since each update needs to write multiple rows in the AMs and is thus not desirable for implementing PER.

\subsection{DQN Execution Latency Analysis}\label{sec:breakdown}
To accelerate DQNs, it is important to understand the latency distribution of different operations in a DQN. For the DQN agent shown in Fig.~\ref{fig:DQN}, 
at each time step, the following operations are done: store (storing a transition to the ER memory), ER operation (sampling a batch of transitions), train (training the target network), and action (action network inference to determine the action to take). Note that PER needs to update the priorities of the sampled transitions, which is also included in the latency of ER operation. We profile the DQN agent on the CartPole and Atari Pong environments running on a NVIDIA GTX 1080 GPU. Two kinds of ER are considered: uniform ER (UER) and PER. The network architectures are the same as in~\cite{mnih2013playing}, i.e., a 3-layer MLP for the CartPole environment, and a 3-layer CNN for the Atari Pong environment. Usually, the ER memory size (the number of experiences) for a complex environment is set to $10^6$ experiences. To study the relationship between the ER memory size and operation latency breakdown, we vary the size of ER memory. Furthermore, we consider two different total numbers of time steps. 

\begin{figure}[t]
    \centering
    \includegraphics[width=0.9\linewidth]{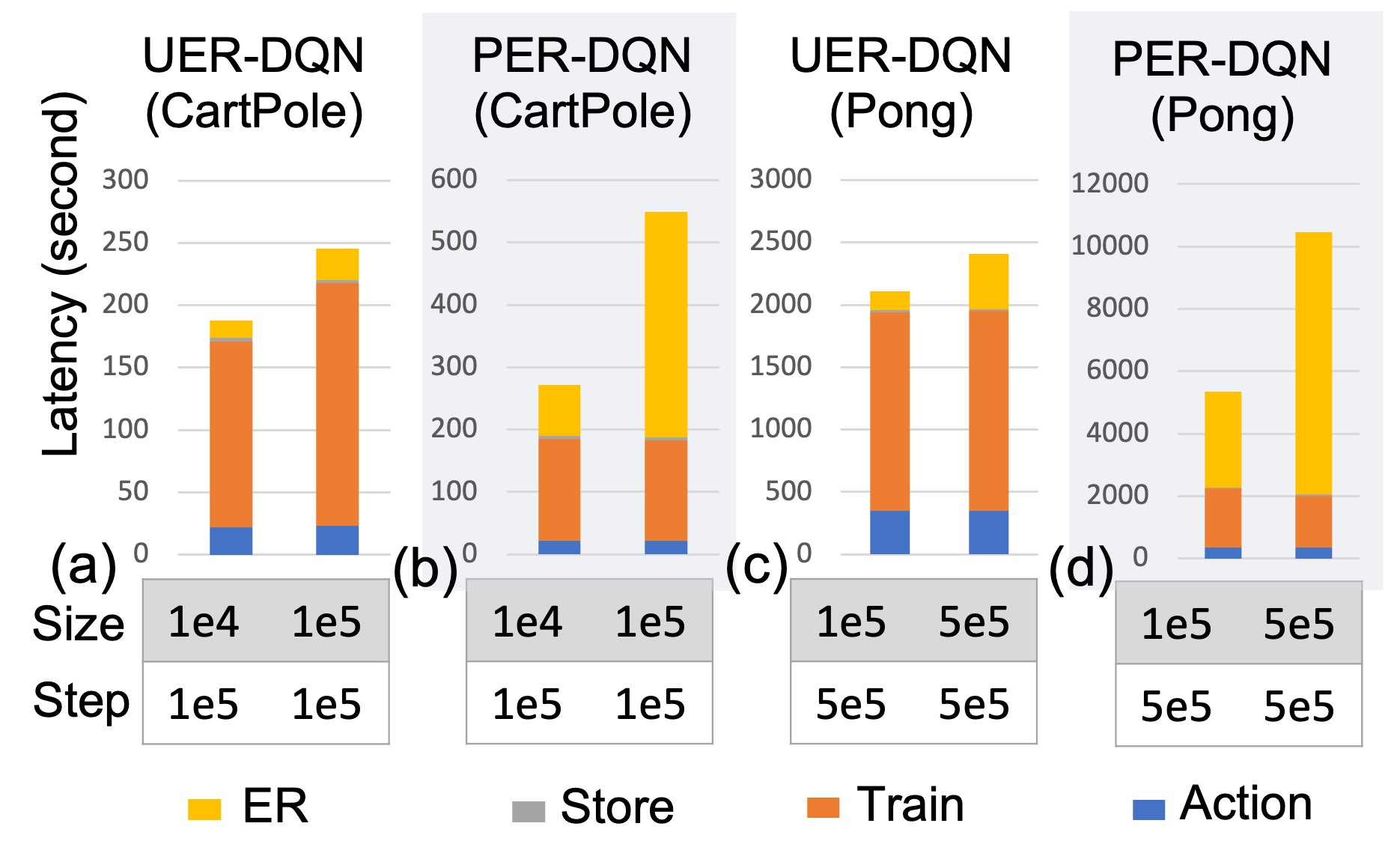}
    \caption{Latency breakdown for executing the UER-DQN and PER-DQN algorithm for the CartPole and Atari Pong environment. Size is the ER memory size and step is the total number of time steps.}
    \label{fig:Breakdown}
    \vspace{-3ex}
\end{figure}

Fig.~\ref{fig:Breakdown} summarizes the profiling results with the corresponding ER memory size and the number of time steps. We observe several trends regarding the ER techniques and the ER memory size when comparing Fig.~\ref{fig:Breakdown}(a) with Fig.~\ref{fig:Breakdown}(b) and Fig.~\ref{fig:Breakdown}(c) with Fig.~\ref{fig:Breakdown}(d). First, ER operations in PER take much more time than in uniform ER. The reason is that despite the uniform random number generation process, sampling operation in PER (discussed in Sec.~\ref{sec:PER}) needs to search on the sum tree structure and the tree needs to be updated with new priority values, which incur many tree-traversal steps.
Second, a larger ER memory size can result in even longer time spent in ER over training due to the deeper tree depth. Third, when the ER memory size increases to $10^5$, the ER operation takes nearly 50\% of the total operation time. According to a study on the size of the ER memory~\cite{fedus2020revisiting}, it is necessary to have a large ER memory to improve the learning performance (i.e., the global return) of the agent. Thus, in the state-of-the-art DQN, PER is a bottleneck in accelerating the learning process.

\section{Associative Memory based PER} \label{sec:imrs}


This section presents our hardware-software co-design approach to accelerate PER. On the software side (Sec.~\ref{sec:overview}--\ref{sec:cnn}), we introduce a novel algorithm \fw which leverages AM-friendly priority sampling operations to approximate the original priority sampling technique. On the hardware side (Sec.~\ref{sec:hardware}), we design a AM-based in-memory computing architecture to support \fw efficiently with fast search and update. 

\subsection{Overview of \fw}\label{sec:overview}

As discussed in Sec.~\ref{sec:breakdown}, PER faces memory access challenges due to frequent sampling and update operations. We aim to introduce an alternative PER method such that it can leverage the power offered by in-memory computing while preserving the learning performance offered by the original PER. In this subsection, we first present a high-level idea on approximating the priority sampling operation, and then give an overview of \fw.

Intuitively, priority sampling aims to sample a higher-priority experience with a higher probability. Fig.~\ref{fig:PERIdea}(d) illustrates a straightforward way to transform priority sampling to uniform sampling. Here, we store multiple copies of the same priority, where the number of copies corresponds to the magnitude of the priority value. For example, we store three copies of $p_{1}$, two copies of $p_2$, etc., and a total of 11 entries. Now if we uniformly sample the 11 entries, the probability of sampling $p_{1}$ is $p_{1}/S$. It is easy to see that the sampling speed of this method should be much faster than the tree-based solution (Fig.~\ref{fig:PERIdea}(c)). However, the method would require a huge amount of memory, especially when the priority values are large. 

Inspired by the idea shown in Fig.~\ref{fig:PERIdea}(d), we develop \fw by using uniform sampling while minimizing memory requirements for storing priorities. Specifically, we propose to construct a subset of the priorities for uniform sampling such that the count of large priorities is higher than that of small priorities. The subset is referred to as the candidate set of priorities (\csp). Now if we uniformly sample the \csp, the larger priorities will be selected with higher probabilities. A key question then is how to construct the \csp so that the final learning performance would not be degraded.

%
%

To constructs \csp in \fw we first divide all priorities into $m$ groups, where $m$ is a hyper parameter and bears some similarity to quantization level. Given the range of priority values as $[0, V_{max}]$, group $g_{i}$ represents the value range $[\frac{V_{max}*i}{m}, \frac{V_{max}*(i+1)}{m}]$, where $g_{0} \cup g_{1} \cup \cdot \cdot \cdot \cup g_{m-1} = [0, V_{max}]$. For group $g_{i}$, the count of priorities in $g_{i}$ is denoted by $C(g_{i})$.

\begin{figure}[t]
    \centering
    \includegraphics[width=\linewidth]{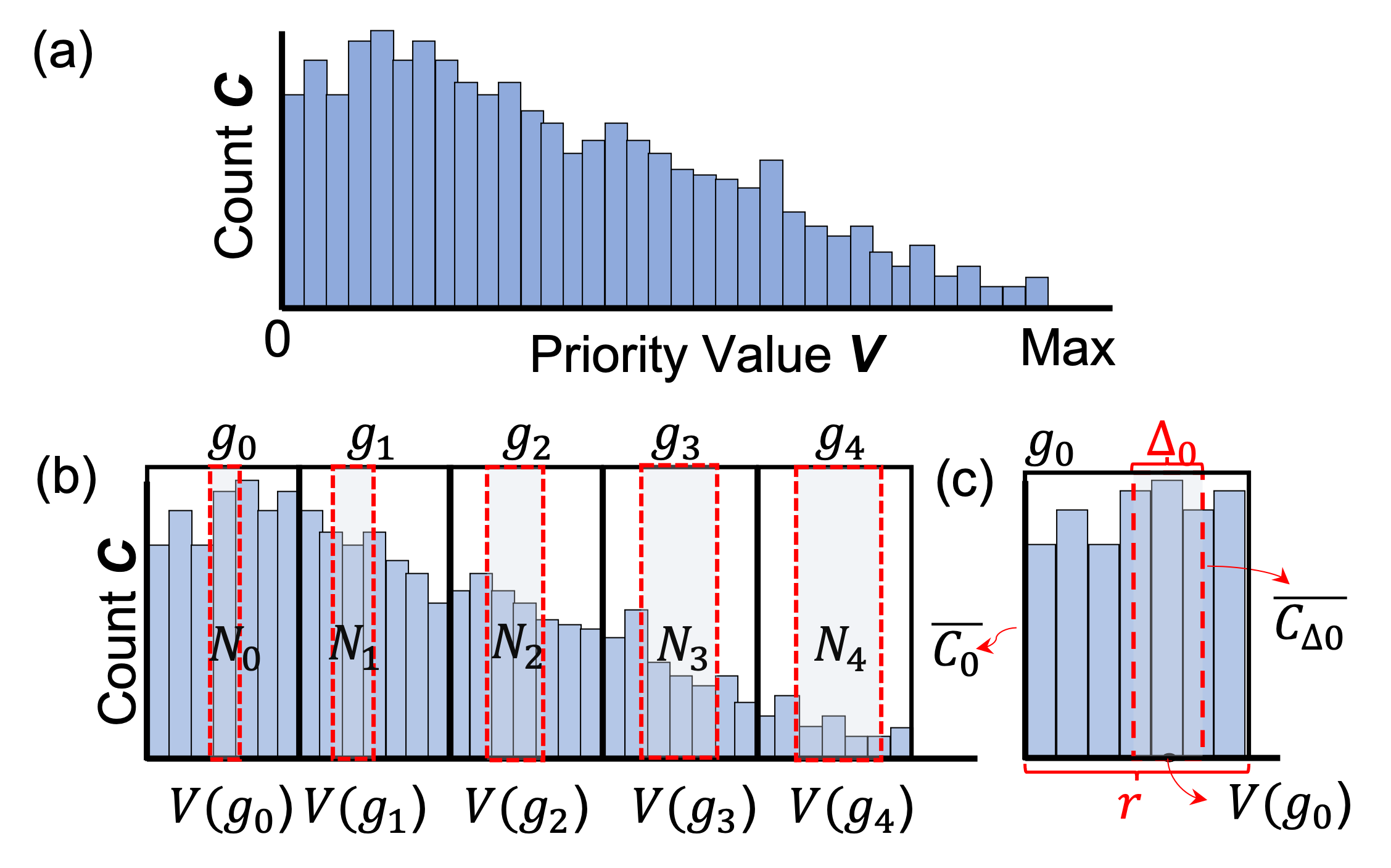}
    \caption{Key \fw concepts: (a) Distribution of all priorities. X-axis is the priority value. Y-axis is the count corresponding to each distinct priority value. (b) Example of kNN based \fw. 5 ($m=5)$ groups are used (separated by thick black lines), and the priorities in the red-dashed blocks are selected. (c) Example of \fnn based \fw. One group is shown as other groups follow the same idea.}
    \label{fig:cNN}
    \vspace{-2ex}
\end{figure}

Consider the simplest case that we set $m$ equal to the number of distinct priority values. Then, all the priorities within the same group have the same priority value. Fig.~\ref{fig:cNN}(a) depicts a representative distribution of priorities in a DQN, where each vertical bar corresponds to one group. Since the probability to sample the priorities in the same group should be equal, we can simply choose a subset of priorities from every group to form the \csp. If we let the size of the subset for $g_i$ to be proportional to $C(g_{i}) \cdot V(g_{i})$, where $V(g_{i})$ denotes the priority value for group $g_i$, for a larger priority value $V(g_{i})$, more priorities are included in the \csp. 

The key idea behind \fw is thus to approximate priority sampling with uniform sampling by constructing a representative \csp. However, several challenges still exist: (1) though the simplest method of setting $m$ is straightforward and incurs little learning performance loss, the \csp can still be very large since the range of priority values is usually large. (2) Selecting priorities based on the value magnitude in run-time requires the priority list always sorted, which is costly to implement in CPU/GPU.

\subsection{Nearest Neighbor \fw}\label{sec:knn} 

In this section, we present a k-Nearest Neighbor(kNN) based priority sampling method to tackle the two challenges discussed above. This method would be able to exploit efficient search provided in AM.
First, we set hyper parameter $m$ much smaller than the number of distinct priority values to reduce the \csp size. 
In group $g_i$ (for $0 \leq i \leq m-1$), the priority values are within the range of $[\frac{V_{max}*i}{m}, \frac{V_{max}*(i+1)}{m}]$ and the number of all the priorities with values within this range is $C(g_{i})$. Following the idea discussed in Sec.~\ref{sec:overview}, we need to determine the representative priority $V(g_{i})$ and the subset size of $g_{i}$ to construct \csp.
For representative priority value $V(g_{i})$, we randomly select a priority value in $[\frac{V_{max}*i}{m}, \frac{V_{max}*(i+1)}{m}]$, which preserves the randomness requirement in PER. We denote the subset size of $g_i$ as $N_{i}$ and set it as
\begin{equation}
\label{N}
    N_{i} = \lambda \cdot V(g_{i}) \cdot C(g_{i}),  i\in[0, m-1],
\end{equation}
where $\lambda$ is a scaling factor that scales the subset size and linearly correlates with the \csp size. $\lambda$ is another hyper parameter which can be tuned to trade off learning performance and hardware cost. (The impacts of $\lambda$ and $m$ will be studied in Sec.~\ref{sec:algorithm}).


Second, a kNN search process is employed to construct \csp with a few search steps without keeping the priority list sorted. Specifically, to obtain the subset of $g_i$, we choose $N_{i}$ priorities with values closest to $V(g_{i})$, which can leverage efficient search supported by AM. The rationale is that these priority values are good representatives for $g_i$. Note that simply randomly picking $N_i$ values from $g_i$ would be more expensive to implement in hardware and is thus avoided. The \csp is the union of the subset of each group as illustrated in Fig.~\ref{fig:cNN}(b). We uniformly sample from the resulting \csp to get the sampling result.  Algorithm~\ref{alg:gradient_gating} (ignoring Line 9--12 for now) summarizes the \fw method described above, which is referred to as \textbf{\fwk}. It can be seen that kNN search is the main operation in this \fw implementation to find all $N_i$ candidates. 

\subsection{Approximate Nearest Neighbor \fw}\label{sec:cnn}
As shown in Algorithm~\ref{alg:gradient_gating}, the main operation in \fwk is kNN search. But two facts may limit the adoption of the method in an AM based architecture. First, the search function in AM requires a different sensing circuit than traditional AMs~\cite{hu2021memory}. Typically, the sensing circuit for NN search incurs additional latency and area cost~\cite{kazemi2021memory}. Also, several (k) search operations are needed to find all (k) neighbors. Second, to ensure we obtain $N_{i}$ priorities for each group $g_i$, we need to keep track of the total priority count in each group, which requires additional circuitry. Below we introduce another variant of \fw, which approximates kNN search with fixed-radius Nearest Neighbor search (\fnn), \fwf.

Fixed-radius nearest neighbor search, also known as C-Nearest Neighbor search, finds neighbors of the query within distance $C$. Fig.~\ref{fig:cNN}(c) illustrates the concept of \fwf. The key idea is to employ parameter $\Delta_{i}$ representing the distance from value $V(g_{i})$. 
Then, we use \fnn search, find all the neighbors of $V(g_{i})$ within distance $\Delta_{i}$, and obtain a subset of $g_i$. Similar to \fwk, \fwf constructs the \csp by taking the union of the subsets of all $g_i$'s. Now, if we apply uniform sampling on the resulting \csp, we expect the sampled priority to be similar to that obtained by PER. 

To ensure such a similarity holds, the key is determining an appropriate $\Delta_{i}$. We derive the appropriate $\Delta_{i}$ according to Eqns.~\eqref{Delta}-~\eqref{delta}.
Given distance $\Delta_{i}$, the number of all priorities inside the range is $C_{\Delta i}$, which is expected to be an approximation of $N_i$ in ~\eqref{N}. Thus, the average number of all the distinct priority values within $\Delta_{i}$ is

\begin{equation}
\label{Delta}
\overline{C_{\Delta i}} =\frac{C_{\Delta i}}{\Delta_{i}}   \approx \frac{N_i}{\Delta_{i}},
\end{equation}
where $\Delta_{i}$ can be determined by $\overline{C_{\Delta i}}$ and $N_i$. However, the challenge with this approach is that it is difficult to obtain the exact value of $\overline{C_{\Delta i}}$ because the distribution of the priorities changes from time to time as new experiences are put into ER memory. To address this issue, we propose to use the average number of all the distinct priority values within group $g_i$, $\overline{C_{i}}$, to approximate $\overline{C_{\Delta i}}$ as
\begin{equation}
\label{AverageC}
    \overline{C_{\Delta i}} \approx \overline{C_{i}} = C(g_{i})/r,
\end{equation}
where $r$ is the group range size. From Eqn.~\eqref{N}, ~\eqref{Delta}, and~\eqref{AverageC}, we have

\begin{equation}
\label{delta}
    \Delta_{i} \approx  \frac{N_i \cdot r}{C(g_i)} = \frac{\lambda \cdot V(g_{i}) \cdot C(g_{i}) \cdot r}{C(g_i)} 
    \\= \lambda \cdot V(g_{i})\cdot \frac{Vmax }{m} =   \frac{\lambda^{’}}{m} \cdot V(g_{i}).
\end{equation}

Based on Eqn.~\ref{delta}, we can calculate $\Delta{i}$ for each group to be used in the \fnn search by only knowing $V(g_{i})$, since $\lambda^{’} \& ~ k$ are hyper parameters. Thus, in \fwf, for the $i$-th group, we search for neighbors of $V(g_{i})$ within $\Delta{i}$ distance (Fig.~\ref{fig:cNN}(c)) to construct the \csp. Algorithm~\ref{alg:gradient_gating} (ignoring lines 4--8) summarizes the sampling process in \fwf. This design enables a faster AM search process and avoids the overhead of tracking priority counts in each group.

\begin{algorithm}[t]
\SetAlgoLined
\KwIn{All priorities $p$, group number $m$, scaling factors $\lambda$, $\lambda^{'}$, maximum priority value $V_{max}$, batch size $b$}
\KwOut{Sampled priority set in $sp$}
$CSP$ = []\;
\For{$i$ in range($m$)}{
    $V(g_{i})$ = random.uniform($\frac{V_{max}}{m}\cdot i$, $\frac{V_{max}}{m}\cdot (i+1)$) \;
    
    \If{kNN}{
       $C(g_{i})$ = count($p_{i}$) in range($\frac{V_{max}}{m}\cdot i$, $\frac{V_{max}}{m}\cdot (i+1)$)\;
    
    $N_{i}$ = round($\lambda \cdot V(g_{i}) \cdot C(g_{i})$)\;
    $CSP$.add(kNN($V(g_{i}), N_{i}$))\;
    }\ElseIf{frNN}{
    $\Delta_{i}$ = round($\lambda^{'} / m \cdot V(g_{i})$)\; 
    $CSP$.add(frNN($V(g_{i}), \Delta_{i}$)) \;
    }
}
\For{$j$ in range($b$)}{
$id$ = random.uniform(len($CSP$))\;
    $sp$.add($CSP$[$id$]);\
    }
\Return  $sp$

\caption{\fw. (The {\tt if} condition at lines 4--8 are for kNN variant and lines 9--12 describe the \fnn variant.)  }\label{alg:gradient_gating}
\end{algorithm} 
 
\subsection{Hardware Support for \fw}\label{sec:hardware}
Here we present a hardware design to support \fw. We elaborate on the high-level architecture and the operation flow, and provide details of the search methodology and additional circuits. 

\begin{figure}[t]
    \centering
    \includegraphics[width=\linewidth]{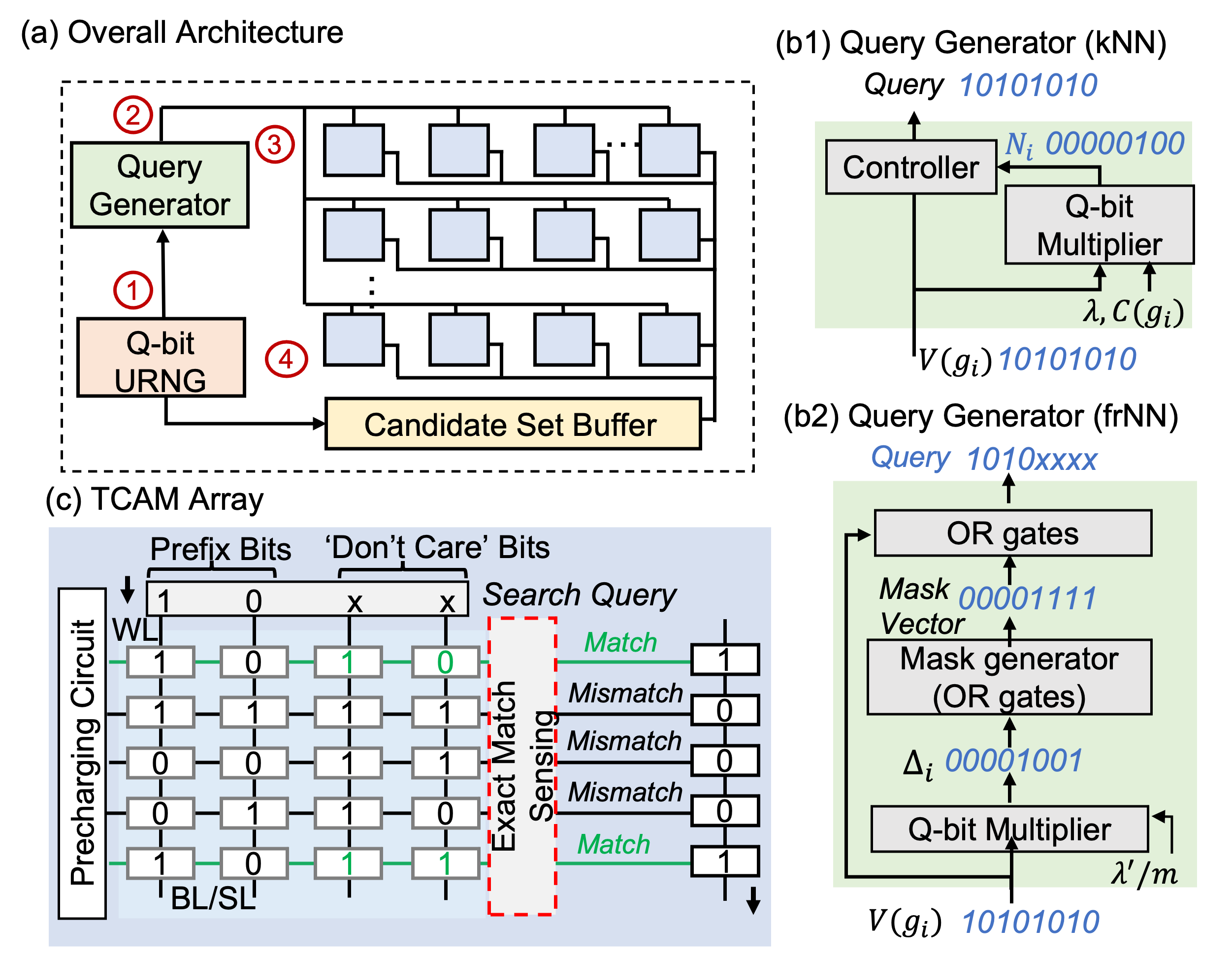}
    \caption{(a) Overview of the AM based architecture for supporting \fw. Each blue-shaded box represents a TCAM array. (b1) Query generator design for \fwk. (b2) Query generator design for \fwf with the prefix-based query generation. Query generation examples (Q=8 bits) are shown in (b1)(b2). (c) TCAM array with prefix-based query strategy and exact match sensing used for \fwf.}
    \label{fig:hardware}
    \vspace{-2ex}
\end{figure}

The AM-based \fw accelerator architecture is shown in Fig.~\ref{fig:hardware}(a). The design supports parallel search with multiple TCAM arrays and contains a uniform random number generator (URNG), a query generator, and a candidate set buffer. The architecture works as follows: (1) The URNG generates a random search query $V(g_{i})$ for each group (line 3 in Algorithm~\ref{alg:gradient_gating}). (2) The query generator generates the corresponding search query for each group, and the query is sent to all TCAM arrays (lines 6\&10 in Algorithm~\ref{alg:gradient_gating}). (3) Multiple TCAM arrays work in parallel to find all matching entries, which are sent to the candidate set buffer (lines 7\&11 in Algorithm~\ref{alg:gradient_gating}). (4) In the last step, the URNG generates a batch of random numbers, and the corresponding entries in the candidate set buffer are accessed and used as the final output (lines 15\&16 in Algorithm~\ref{alg:gradient_gating}). For both the \fw variants, the same dataflow can be deployed with minor differences in the query generator and the sensing circuit of the TCAM array, which are introduced below.

\subsubsection{\fwk Search}
The query generator for the kNN variant shown in Fig.~\ref{fig:hardware}(b1), implements Equ.~\ref{N} to calculate the expected CSP size $N_{i}$ by using a $Q$-bit multiplier. The $Q$-bit input $V(g_{i})$ will be output as the search query multiple times. The search time is controlled by $N_{i}$. As reviewed in Sec.~\ref{sec:am}, TCAM supports searching all the data stored in the TCAM array in one step. To realize the kNN search in AM, TCAM arrays with \textbf{best match sensing circuit}~\cite{iedm2021} can be deployed (red dotted block in Fig.~\ref{fig:hardware}(c)). For each search operation, the neighbor nearest to $V(g_{i})$ is output, and multiple search operations are needed to find $N_{i}$ nearest neighbors. Besides the multiple search steps needed, there are other challenges with the kNN implementation. Best match search requires more sophisticated sensing circuitry since an accurate comparison of the number of matching cells is needed. Furthermore, the search accuracy can suffer significantly when the number of cells in a row is large and there are non-negligible device variations and noises~\cite{kazemi2021fefet}. 

\subsubsection{\fwf Search}
For \fwf search, we devise a \textbf{prefix-based query strategy}, an efficient query mapping technique, which approximates the search radius by using the bit properties of fixed-point values. This approach only requires \textbf{exact-match} TCAMs which employ very simple sensing circuitry since only match or mismatch need to be differentiated. Furthermore, only one search operation is needed to get all candidates. The prefix-based query strategy follows the steps below to select all neighbors within $\Delta_{i}$ distance of $V(g_{i})$. 

First, the query generator is designed as shown in Fig.~\ref{fig:hardware}(b2), which consists of a $Q$-bit multiplier, a mask generator, and $Q$ OR gates. A three-step prefix generation works as follows: 1) The multiplier generates search range $\Delta_{i}$ following Equ.~\ref{delta}. 2) The mask generator finds the position of the leftmost `1' in $\Delta_{i}$, called `p', which determines the position of prefix bits and don't care bits in the mask vector. In the mask vector, all bits to the left of `p' are set to `0' and all bits to the right of `p' (including `p') are set to `1'. The mask generator is implemented using OR gates. 3) Given the input $V_{gi}$ and mask vector of $\Delta_{i}$, the OR gates generate a query composed of prefix bits and don’t care bits. An 8-bit (Q=8) prefix query generation example is shown in Fig.~\ref{fig:hardware}(b2) where `p' is 4. By employing the TCAMs, all the rows that match the query are identified. For the example in Fig.~\ref{fig:hardware}(c), query 10xx will match with the entries within the range (1000,1011). Thus, the number of don't care bits in the query corresponds with the size of the search range. Note that this prefix-based mapping does introduce some approximation error when $V(g_{i})$ and $\Delta_{i}$ are not powers of 2 since the accepted range can only be powers of 2. Detailed latency comparison between the two variants' implementation will be presented in Sec.~\ref{sec:latency}.


\subsubsection{Update in \fw}
As we mentioned in Sec.~\ref{sec:PER}, the sum tree in original PER implementation is updated when updating the priority value (leaf node), which also incurs lots of tree-traversal steps. However, the priority update operation in the proposed \fw is relatively simple because each priority has only one copy in the ER memory. To update the priority, we write the new priority value in AM directly using the write port of TCAM arrays, which is also much faster than the original PER.


\section{Evaluation}
\label{sec:eva}


We evaluate the proposed \fw design with respect to algorithm-level performance and execution latency. 
We first present the algorithm-level performance study in Sec.~\ref{sec:algorithm}. Then array-level and end-to-end latency study is discussed in Sec.~\ref{sec:latency}.

\subsection{Algorithm-level Performance Study}\label{sec:algorithm}
\subsubsection{Sampling Error Study}

\begin{figure}[t]
    \centering
    \includegraphics[width=\linewidth]{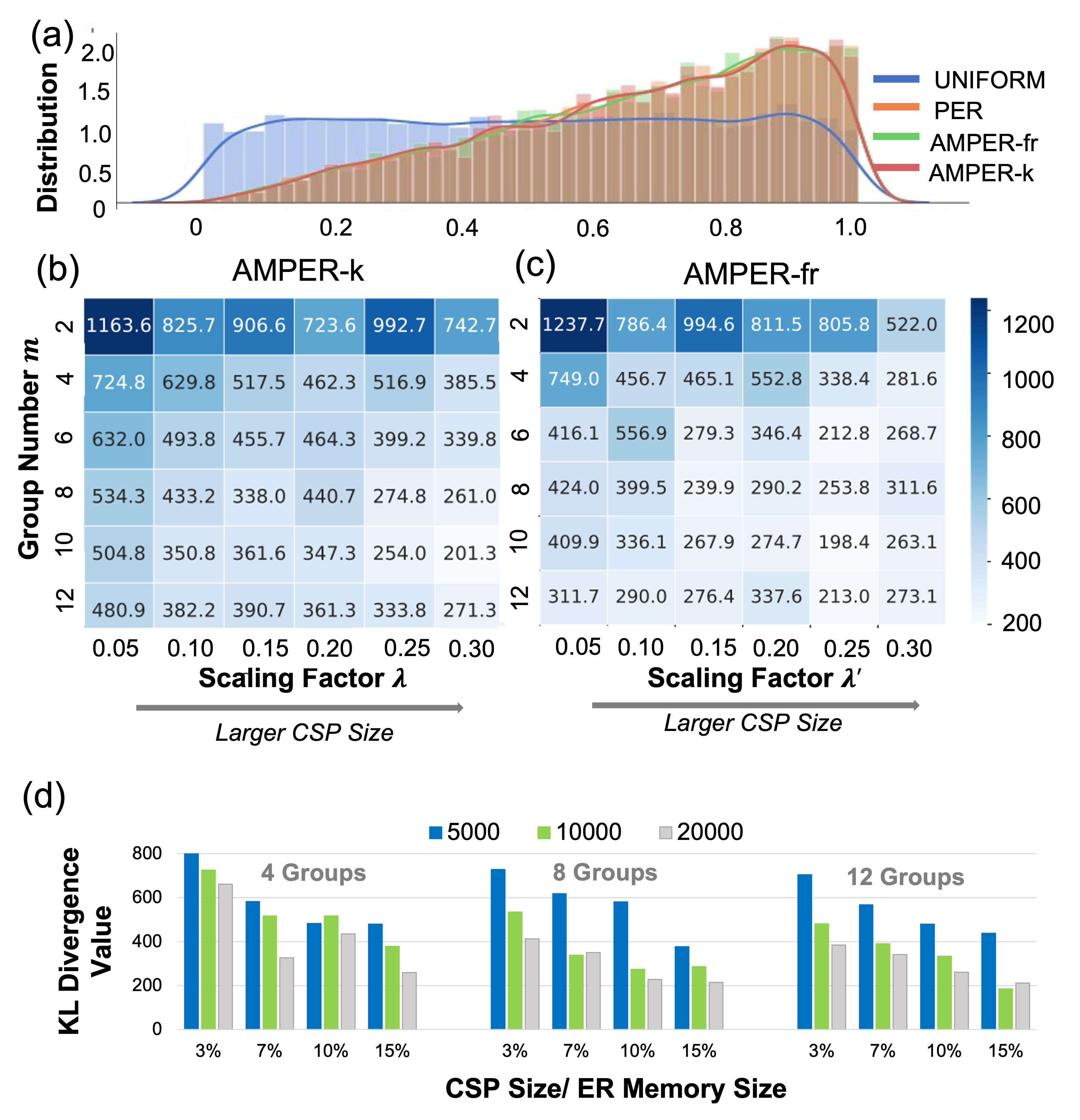}   
    \caption{Sampling error study. (a)Visualization of the sampling results from Uniform, \fwk, \fwf and PER methods. KL Divergence study of (b) \fwk (c) \fwf under different group numbers and scaling factors. (d) KL Divergence study  of \fwk for different ER memory sizes (5000, 10000, 20000).}
    \label{fig:KL}
    \vspace{-2ex}
\end{figure}

Since \fw adopts a novel AM-friendly priority sampling concept, it is important to compare \fw and PER regarding the sampling performance. Specifically, we compare the sampling results from PER and \fw. First, we generate a random data list with size 10000 from an uniform distribution within the range [0,1] and sample it with PER and \fw, respectively. The sampling is repeated with batch size 64 for 100 runs. The sampling results distribution is visualized in Fig.~\ref{fig:KL}(a), where two variants of \fw both generate similar results as the standard PER with the curve mostly overlapped.

To further analyze the sampling difference between \fw and PER, we quantify the difference using the metric Kullback-Leibler (KL) Divergence, a measure of how one probability distribution is different from another reference probability distribution with the unit `nat'. Smaller KL Divergence values indicate more similar distributions. Given two discrete distributions $P, Q$, the KL divergence is defined as 
$KL(P,Q):= SUM(P[i] * log(P[i] / Q[i] ), i).$

As we discussed in Sec.~\ref{sec:knn}, the scaling factor $\lambda$ and group number $m$ are two key hyper parameters in \csp construction. We vary the values of $\lambda$ and $m$ and repeat the sampling process with batch size 64 for 100 runs to generate different sampling results. Fig.~\ref{fig:KL}(b)(c) shows the KL divergence values between the AMPER and the PER sampling results under different hyper parameter combinations. Group number $m$ is shown along the y-axis for the two figures, increasing from 2 to 12. X-axis depicts the scaling factor $\lambda$/ $\lambda'$ which linearly correlates with the size of the \csp. As shown in Fig.~\ref{fig:KL}(b) and (c), for both \fw variants, increasing group number $m$ and scaling factor $\lambda$/ $\lambda'$ decreases the KL Divergence value, which means less sampling error. AMPER introduces a large sampling error when the group number and scaling factor are very small (upper left corner), say $m=2, \lambda=0.05$. However, at the bottom right corner that AMPER has less than 300 nats KL divergence, which is quite similar to the original PER. For reference, the KL Divergence value between uniform sampling and PER sampling is around 9000 nats, and the KL Divergence value between different runs of PER is around 140 nats. Thus, by choosing proper hyper parameters, AMPER can achieve a similar sampling performance as PER. Comparing Fig.~\ref{fig:KL}(b) and (c), \fwf also achieves comparable performance as \fwk. Later in Sec.~\ref{sec:latency}, we will further discuss the impact of hyper parameters on execution latency. 

Fig.~\ref{fig:KL}(d) studies the sampling error under different ER memory size. We vary the ER memory size from 5000 to 20000 for \fwk. For each ER memory size, the group number is set to 4/8/12, and the x-axis is CSP ratio (CSP size / ER memory size). Fig.~\ref{fig:KL}(d) shows that the findings in Fig.~\ref{fig:KL}(b)(c) still hold for different ER memory sizes. Also, under the same $m$ and the CSP ratio, \fw achieves better sampling performance when the ER memory size becomes larger. The same trends hold for \fwf.




\subsubsection{DQN Learning Performance Study}
To study the performance of \fw in DQN learning, we implemented \fw using PyTorch and tested it on the learning environments CartPole, Acrobot, and LunarLander provided by OpenAI Gym~\cite{1606.01540}. The action/target networks and their basic hyper parameters are set as~\cite{hessel2018rainbow}. We fix the number of steps for each environment. If the ER memory is full, it discards the oldest experience. The training score is the return of each training episode, and the test score is the average return of 10 episodes. The return is defined as the accumulated reward over an episode. Each environment defines its own rewards. For Cartpole, +1 reward is given at a timestep if the pole remains upright. The Acrobot environment gives a reward of -1 at each time step before the problem is solved. For LunarLander, the environment gives either positive or negative rewards at each step. The higher the score, the better the agent learning performance.

\begin{figure}[t]
    \centering
    \includegraphics[width=\linewidth]{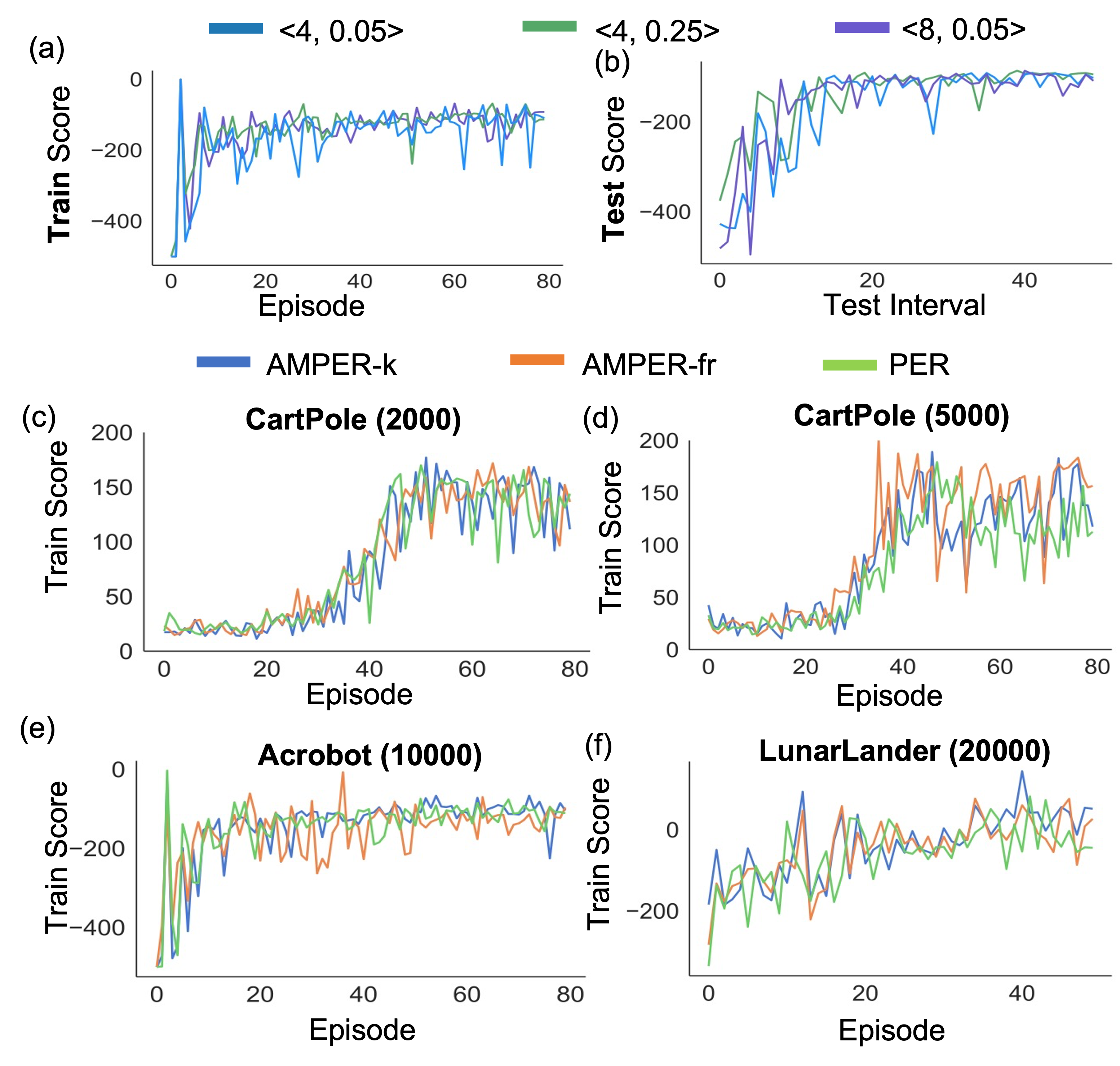}
    \caption{(a) train scores and (b) test scores when training with different hyper parameters for \fwk on the Acrobot environment. Group size and scaling factor are set to <4, 0.05>, <4, 0.25>, <8, 0.05>, respectively.
    Training scores of the DQN agent using PER, \fwk, and \fwf with different ER memory sizes: (c) CartPole with size 2000; (d) CartPole with size 5000; (e) Acrobot with size 10000; (e) LunarLander with size 20000. The scores are averaged over 3 runs.}
    \label{fig:Training}
\end{figure}

We first evaluate the relationship between the sampling error and DQN learning performance. We choose three sets of <$m, \lambda$> combination: <4, 0.05> / <4, 0.25> / <8, 0.05>, which correspond to KL divergence value of 724.8 / 516.9 /534.3 nats, respectively. Fig.~\ref{fig:Training}(a)(b) show the training score and test score curves of the DQN agent on the Acrobot environment with ER memory size 10000. It can be seen that the <4, 0.05> (blue curve) combination, which has the largest sampling error, learns slowest and has the most unstable training curve compared with the other two that exhibit similar learning performance. However, the three settings still reaches to similar final score (Fig.~\ref{fig:Training}(b)) and \fw works well in DQN learning even with large sampling error (around 700 nats).

Fig.~\ref{fig:Training}(c)-(e) show the learning curves of the DQN agent using PER and the proposed \fw on the three environments. \fwf and \fwk exhibit comparable learning speed and final score as PER. Table~\ref{tab:score} summarizes the test scores for the different tasks and ERM sizes. \fw achieves little score degradation compared with PER. Moreover, \fwk achieves even better performance in some cases (CartPole-2000, Acrobot, LunarLander). 

Comparing \fwk and \fwf, the kNN variant has a more stable learning process and better final score than the \fnn variant. The reason is that Eqn.~\eqref{AverageC} incurs approximation errors in the \fnn implementation. If the average count of the selected subset (i.e., $\overline{C_{\Delta i}}$) is quite different from the average group count ($\overline{C_i}$), an error is introduced when calculating $\Delta_i$. However, \fwf still exhibits similar performance as PER in most cases, and provides much faster speed to be shown in Sec.~\ref{sec:latency}.

\subsection{Hardware Performance Study}\label{sec:latency}

   

\subsubsection{Experimental Setup}
To evaluate the proposed hardware accelerator for \fw, we developed RTL-level Verilog models for URNG circuits and the query generator (QG) and synthesized them with Cadence Encounter for a CMOS 45nm library\footnote{We used 45nm CMOS technology library in order to be consistent with the TCAM data reported in~\cite{ni2019ferroelectric}.}. Each priority entry is represented with INT-32 bits. The bit-width Q is set to 32 for each component. The URNG is implemented with the 32-bit linear feedback shift register. The latency consumed by the  candidate set buffer are calculated using CACTI~\cite{balasubramonian2017cacti}. A candidate set buffer (CSB) with a size of 0.3MB is used, which can hold 8000 entries in total. We employ the CMOS-based 16T TCAM design with the best match~\cite{iedm2021} and exact match sensing circuits~\cite{ni2019ferroelectric} for the proposed hardware accelerator. Each TCAM array is 64 rows $\times$ 64 columns, where each row stores a priority entry. Multiple arrays (e.g. 128 arrays for ER memory size 8,192) are needed to store all the priorities. Table.~\ref{tab:latency} summarizes the latency of each component. The latency of PER is measured on the system with Intel i5-8600k CPU and Nvidia RTX 1080 GPU. Sampling is done in batches of 64, that is, 64 priorities are returned after each sampling.

\begin{table}[t]
    \centering
    \caption{Test score comparison of PER, \fwk and \fwf on the OpenAI environments (CartPole, Acrobot, LunarLander).}
    \begin{tabular}{c|c|c|c|c}
    \hline
    \hline
        \textbf{Env} & \textbf{Size} & \textbf{PER} & \textbf{\fwk} & \textbf{\fwf}\\
    \hline
    \hline
    CartPole & 2000 & 162.20 & \textbf{180.13} & 154.18\\
    \hline
    CartPole & 5000 & \textbf{177.32} & 173.20 & 173.25  \\
    \hline
    Acrobot & 10000 & -89.39 & \textbf{-88.89} & -93.69 \\
    \hline 
    LunarLander & 20000 & 185.33 & \textbf{200.10} & 161.50 \\
      
    \hline
    \hline
    \end{tabular}
    \label{tab:score}
\end{table}

\begin{table}[t]
    \centering
    \caption{Latency of \fw hardware components.}
    \begin{tabular}{c|c|c|c}
    \hline
    \hline
    Component &\multirow{2}{*}{\begin{tabular}[c]{@{}c@{}}TCAM Array\\ (Exact~\cite{ni2019ferroelectric})\end{tabular}}
    &\multirow{2}{*}{\begin{tabular}[c]{@{}c@{}}TCAM Array\\ (Best~\cite{iedm2021}) \end{tabular}} & \multirow{2}{*}{\begin{tabular}[c]{@{}c@{}}CSB \\(0.03MB) \end{tabular}} \\
    &&\\
    \hline
    Operation& Search/Write&Search/Write&Read / Write \\
    \hline
    \textbf{Delay (ns)}&0.58 / 2.0  & 1.0  / 2.0  & 0.78 / 0.78  \\
    \hline
    \hline
    Component & URNG  & QG (kNN) &QG  (\fnn)\\
    \hline
    \textbf{Delay (ns)} &1.71 & 3.57 & 2.02 \\
    \hline
    \hline
        
    \end{tabular}
    \label{tab:latency}
    \vspace{-2ex}
\end{table}



\subsubsection{Latency Evaluations}
\begin{figure}[t]
    \centering
    \includegraphics[width=\linewidth]{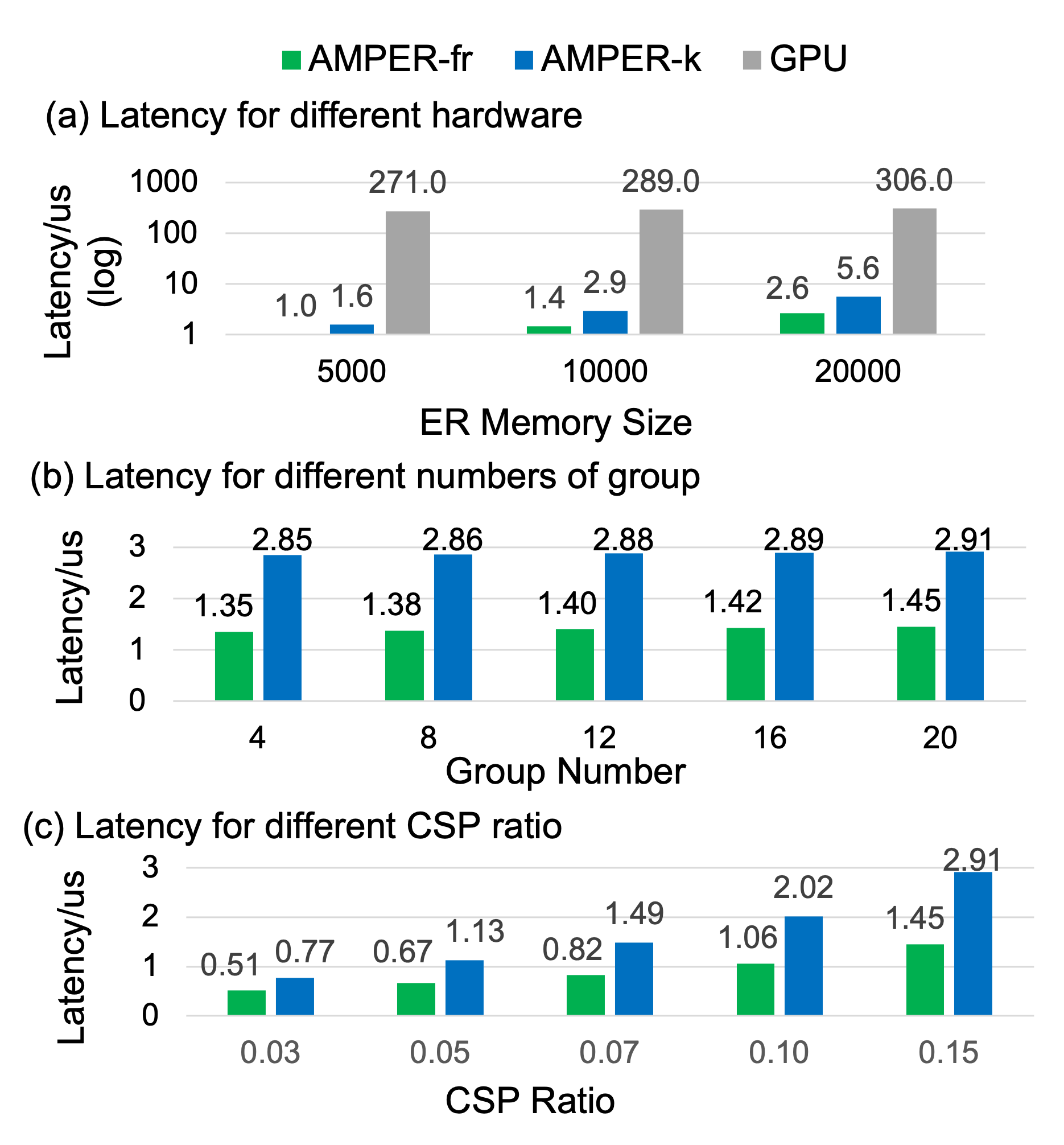}
    \caption{End-to-end latency for \fwf and \fwk. (a) Comparsion with GPU implementation. (b) With CSP ratio at 0.15. Varying group number $m$ from 4 to 20. (c) With group number $m$ at 20. Varying CSP ratio from 0.03 to 0.15.}
    \vspace{-2ex}
    \label{fig:latency}
\end{figure}

We first compare the performance of the proposed accelerator with the GPU implementation. The latency is measured for per batch sampling. To ensure the best learning performance, we set $m$ to 20 and the CSP ratio to 15\%. Fig.~\ref{fig:latency}(a) summarizes the comparison for ER memory sizes 5000, 10000, and 20000. \fwk and \fwf are 55$\times$--170$\times$ and 118$\times$--270$\times$ faster than the GPU implementation, respectively. Note that \fw runs slower on GPU/CPU than the original PER because the nearest neighbor search operation is time-consuming on GPU/CPU. However, our hardware-software co-design approach achieves significant latency improvements over PER. Moreover, based on the data shown in Fig.~\ref{fig:latency}(a), \fw-\fnn achieves  \textasciitilde 2$\times$ latency improvement compared to \fwk. Although the parallel TCAM search ability is exploited in both variants, \fwf achieves better performance due to the simple sensing circuit design. According to the data in Table~\ref{tab:latency}, for each search operation, the TCAM array with best match sensing incurs 1.7$\times$ latency compared with the exact match one due to the more complicated sensing circuit. Furthermore, the number of search operations is reduced by using \fnn search compared with kNN search.

Fig.~\ref{fig:latency}(b) investigates the end-to-end latency of \fwf and \fwk with different group numbers, where the CSP size ratio is fixed to 0.15. The ER memory size is set to 10000 for both implementations. For both implementations, increasing group number has a small impact on the latency. This is because the TCAM array search is done in parallel, which is much faster than other components (Table~\ref{tab:latency}), especially the candidate set buffer write operations. Thus the additional search operations introduced by increasing group number has little impact on the end-to-end latency.

The end-to-end latency for \fwf and \fwk for different CSP sizes is studied in Fig.~\ref{fig:latency}(c), where the group number is fixed to 20. Fig.~\ref{fig:latency}(c) shows that the latency of both \fwf and \fwk increases linearly with the CSP size as the latency is now dominated by the candidate set buffer throughput. As discussed in Fig.~\ref{fig:KL}, increasing both the group number and the CSP size helps improve the algorithm-level performance of the two variants. However, according to the data in Fig.~\ref{fig:latency}(b)(c), increasing the group number is a better option as it incurs limited additional latency to get better sampling performance.



\section{Conclusion}
\label{sec:conclusion}
In this paper, we propose a hardware-software codesign approach \fw to accelerate PER in the state-of-the-art DRL agent. \fw employs the AM-based search operation to approximate PER. An in-memory-computing hardware architecture based on AM is designed to support \fw. \fw shows comparable learning performance as the PER and achieves 55$\times$\textasciitilde270$\times$ latency improvement compared with PER implemented on GPU.



\bibliographystyle{my_abbrv.bst}
\bibliography{references}

\end{document}